\documentclass[11pt,preprint2]{aastex}

\usepackage[]{natbib}
\usepackage{graphicx}
\usepackage{amssymb}
\usepackage{color}
\usepackage{ulem}

\newcommand{\etacar}{$\eta$~Car}

\input{macro.inp}

\slugcomment{ver 2.0}

\shorttitle{X-ray Time Variation of \etacar\ in 2009}
\shortauthors{Hamaguchi et al.}

\begin{document}

\title{X-ray Emission from Eta Carinae near Periastron in 2009\\
I: A Two State Solution}

\author{Kenji Hamaguchi\altaffilmark{1,2}, Michael F. Corcoran\altaffilmark{1,3},
Christopher Russell\altaffilmark{4}, Andrew M.T. Pollock\altaffilmark{5},
Theodore R. Gull\altaffilmark{6}, Mairan Teodoro\altaffilmark{6,7}, 
Thomas I. Madura\altaffilmark{6,8}, Augusto Damineli\altaffilmark{9},
Julian M. Pittard\altaffilmark{10}
}

\altaffiltext{1}{CRESST and X-ray Astrophysics Laboratory, NASA/GSFC,
Greenbelt, MD 20771, E-mail: Kenji.Hamaguchi@nasa.gov}
\altaffiltext{2}{Department of Physics, University of Maryland, Baltimore County, 
1000 Hilltop Circle, Baltimore, MD 21250}
\altaffiltext{3}{Universities Space Research Association,
10211 Wincopin Circle, Suite 500, Columbia, MD 21044}
\altaffiltext{4}{Faculty of Engineering, Hokkai-Gakuen University, Toyohira-ku, Sapporo 062-8605, Japan}
\altaffiltext{5}{European Space Agency, XMM-Newton Science Operations Centre, European Space Astronomy Centre, Apartado 50727,
Villafranca del Castillo, 28080 Madrid, Spain}
\altaffiltext{6}{Astrophysics Science Division, NASA Goddard Space Flight Center, Greenbelt, MD 20771}
\altaffiltext{7}{CNPq/Science without Borders Fellow}
\altaffiltext{8}{NASA Postdoctoral Program Fellow}
\altaffiltext{9}{Instituto de Astronomia, Geof\'isica e Ci\^encias Atmosf\'ericas, Universidade de S\~ao Paulo, Rua do Mat\~ao 1226, Cidade Universit\'aria,
S\~ao Paulo 05508-900, Brazil}
\altaffiltext{10}{School of Physics and Astronomy, The University of Leeds, Woodhouse Lane, Leeds LS2 9JT, UK}

\begin{abstract}
X-ray emission from the supermassive binary system \etacar\ declines sharply around periastron.
This X-ray minimum has two distinct phases ---
the lowest flux phase in the first $\sim$3 weeks and a brighter phase thereafter.
In 2009, the \CHANDRA\ X-ray Observatory monitored the first phase five times 
and found the lowest observed flux at $\sim$1.9$\times$10$^{-12}$~\UNITFLUX\ (3$-$8~keV).
The spectral shape changed such that 
the hard band above $\sim$4~keV dropped quickly at the beginning
and the soft band flux gradually decreased to its lowest observed value in $\sim$2~weeks.
The hard band spectrum had begun to recover by that time.
This spectral variation suggests that the shocked gas producing the hottest X-ray gas near the apex of the 
wind-wind collision (WWC) is blocked behind the dense inner wind of the primary star, 
which later occults slightly cooler gas downstream.
Shocked gas previously produced by the system at earlier orbital phases is suggested to produce the faint residual 
X-ray emission seen when the emission near the apex is completely blocked by the primary wind.
The brighter phase is probably caused by the re-appearance of the WWC plasma, whose emissivity significantly declined during the occultation.
We interpret this to mean that the X-ray minimum is produced by a hybrid mechanism of an occultation and a decline in emissivity of the WWC shock.
We constrain timings of superior conjunction and periastron based on these results.
\end{abstract}
\keywords{Stars: individual (\etacar) --- stars: early-type --- stars: winds, outflows ---  
binaries: general --- X-rays: stars}

\section{Introduction}

A significant fraction of evolved massive stars are binaries \citep{Sana2012}.
Strong winds with velocities of hundreds to thousands of \UNITVEL\ from two stars
collide, thermalize to $\sim$10$^{7}$~\DEGREEKELV, and emit X-rays.
The conventional theories of the wind-wind collision (WWC) phenomenon \citep[e.g.,][]{Usov1992b,Pittard2002b} suggest that the X-ray luminosity
depends on the mass loss rates, the wind velocities and the binary separation.
However, similar binary systems often exhibit very different X-ray properties \citep{Gagne2012}.
The relation of the X-ray luminosity and the stellar parameters needs to be tested with observations.

The evolved massive binary systems, $\eta$ Carinae \citep[\etacar,][]{Davidson2012} and WR~140, have
highly eccentric orbits \citep[$e \sim$0.9,][]{Nielsen2007,Monnier2011,Fahed2011}.
Their stellar separations vary by a factor of $\sim$20,
so that these systems are useful for studying the dependence of WWC X-ray activity on stellar separation.
The X-ray observatory, \RXTE\ \citep{Bradt1993}, monitored both binary systems and
measured a strong increase in the X-ray luminosity that is approximately inversely proportional to the binary stellar separation,
as predicted by WWC theory \citep[e.g.,][]{Corcoran2005}.
However, sudden unpredicted declines occur in their X-ray fluxes around periastron.
In particular, the X-ray decline of \etacar\ to the X-ray minimum extending for a period of $\sim$3 months in 1998.0 and 2003.5
cannot be explained by a simple eclipse model \citep{Ishibashi1999,Parkin2011}.

The unexpected X-ray behavior of \etacar\ may betray extreme physical conditions at the end of a supermassive star's life.
The primary star (\etacar\ A) is one of the most massive stars known, with a current mass of $\gtrsim$90 $M_{\odot}$ \citep{Hillier2001},
and is believed to be in the Luminous Blue Variable (LBV) stage.
One hundred and seventy years after a series of great eruptions formed the bipolar Homunculus nebula (HN),
\etacar\ A still drives strong winds with
\Mdot~$\sim$8.5$\times$10$^{-4}$~\UNITSOLARMASSYEAR\ and $v_{wind}~\sim$420~\UNITVEL\ \citep{Groh2012},
whose enormous kinetic luminosity is about 5$\times$10$^{37}$~\UNITLUMI.
The secondary star is not observed directly,
but the WWC X-rays from \KT $\sim$4~keV plasma and the variable UV photoionization of nearby circumstellar clouds \citep{Iping2005, Mehner2010}
suggest it to be an O supergiant or WN star with 
$v_{wind} \sim$3000~\UNITVEL\ and \Mdot $\sim$ 10$^{-5}$~\UNITSOLARMASSYEAR \citep{Pittard2002,VernerE2005a}.
These strong winds may collide in an extreme way at periastron, when the two stars pass within a few AU of each other.

The \XMM\ \citep{Jansen2001} and \CHANDRA\ \citep{Weisskopf2002} observations revealed two distinct states across the 2003.5 X-ray minimum \citep{Hamaguchi2007b}.
In the first state, which we call ``the deep X-ray minimum", the X-ray emission declined to its minimum observed value,
lasting approximately 3~weeks.
In the second state, which we call ``the shallow X-ray minimum", the emission abruptly increased by a factor of $\sim$3 from the deep minimum level, 
and then the hard band flux above $\sim$6~keV did not change at all through the shallow minimum for 1.5 months.
The \RXTE\ flux curve showed virtually identical changes across the 2003.5 minimum compared to the 1998.0 minimum, 
so we considered the change to be repetitive.
However,
something drastically changed in the system in 2009, causing the X-ray minimum to end a month earlier \citep{Corcoran2010}.
This change may be due to a drop in {\etacar}'s mass loss rate or wind terminal velocity,
but that explanation is controversial \citep{Mehner2010,Teodoro2012}.

\begin{deluxetable}{lclcc}
\tablecolumns{5}
\tablewidth{0pc}
\tabletypesize{\scriptsize}
\tablecaption{Logs of the \CHANDRA\ Observations\label{tbl:obslogs}}
\tablehead{
\colhead{Abbreviation}&
\colhead{Sequence ID}&
\multicolumn{2}{c}{Observation Start}&
\colhead{Exposure}
\\ \cline{3-4}
&&\colhead{Date}&\colhead{$\phi_{\rm X}$}&\colhead{(ks)}\\
\colhead{(1)}&\colhead{(2)}&\colhead{(3)}&\colhead{(4)}&\colhead{(5)}
}
\startdata
CXO$_{\rm 090110}$&9933&2009 Jan. 10, 19:09 	&1.997&13.8\\
CXO$_{\rm 090116}$&9934&2009 Jan. 16, 01:27	&2.000&14.2\\
CXO$_{\rm 090122}$&9935&2009 Jan. 22, 06:42	&2.003&13.8\\
CXO$_{\rm 090129}$&9936&2009 Jan. 29, 06:40	&2.006&13.6\\
CXO$_{\rm 090203}$&9937&2009 Feb. 3, 14:26	&2.009&12.5\\
\enddata
\tablecomments{
Col. (1): Abbreviation adopted for each observation. 
Col. (2): Observation identification number of each observation.
Col. (3): Time of the observation start.
Col. (4): The X-ray ephemeris in \citet{Corcoran2005}, $\phi_{\rm X}$ = (JD[observation start] $-$ 2450799.792)/2024.
Col. (5): Exposure time excluding the detector deadtime.
}
\end{deluxetable}

In 2003, \CHANDRA\ and \XMM\ performed four pointed observations during the shallow minimum, 
but both acquired only one observation each during the deep minimum.
Since the X-ray flux dropped below the \RXTE\ sky background during the minimum,
we could not measure with \RXTE\ the flux variation during the deep minimum, 
which is crucial to understand the cause of the X-ray decline.
To better define the X-ray spectrum, we performed 5 \CHANDRA\ observations around the deep minimum of 2009.0.
This paper reports their flux variations and discusses their origin.

\begin{figure}[ht]
\plotone{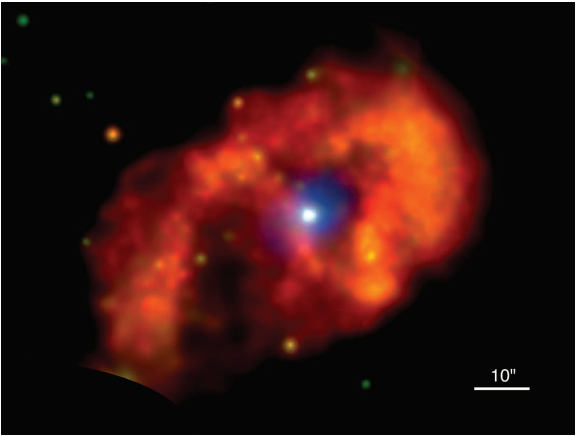}
\begin{scriptsize}
\caption{
True color (red: 0.3$-$1~keV, green: 1$-$3~keV, blue: 3$-$10~keV) image of \etacar\ using all ACIS-S imagery observations 
around the X-ray minimum in 2003 and 2009.
The original image in each color with 0.25\ARCSEC\ pixel bins is smoothed with the adaptive smoothing method ({\tt dmimgadapt}).
\label{fig:obsimg}
}
\end{scriptsize}
\end{figure}

\begin{figure}[hdtp]
\epsscale{0.95}
\plotone{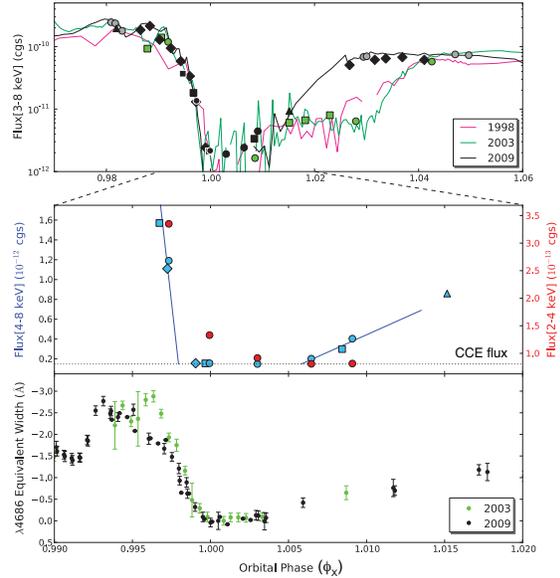}
\vspace{-4mm}
\begin{scriptsize}
\caption{
{\it Top}:
X-ray fluxes (3$-$8~keV) of the CPS (WWC + CCE) measured with \CHANDRA\ (circle), \XMM\ (square), \SUZAKU\ (triangle),
{\it Swift} (diamond) and \RXTE\ (solid lines).
The red, green and black colors represent measurements around the 1998.0, 2003.5 and 2009.0 minima, respectively.
The filled black circles are observed with the \CHANDRA\ ACIS-S in imaging mode, 
while the other circles represent ACIS+HETGS grating observations.
The sky background in the \RXTE\ light curve ($\sim$3$\times$10$^{-11}$~\UNITFLUX)
is estimated based on the \CHANDRA\ observations
during the deep minimum (see Section~\ref{sec:result}) 
and subtracted from the \RXTE\ light curve.
Contribution of the HN emission is also estimated and subtracted from the \XMM, \SUZAKU\ and {\it Swift} data points.
{\it Middle}:
Soft (2$-$4~keV, {\it red}) and hard (4$-$8~keV, {\it blue}) band fluxes of the CPS during the 2009.0 X-ray minimum.
The dotted line is the CCE flux level in both the soft and hard energy bands.
The solid lines show linear extrapolations of 4 hard band data points (1 \SWIFT\ measurement is above this plot) before the ingress
and 3 hard band data points after the egress.
{\it Bottom}:
Variation of the equivalent width of the HeII $\lambda$4686 line \citep{Teodoro2012}.
The color code is the same as in the top panel.
\label{fig:obstiming}
}
\end{scriptsize}
\end{figure}

\section{Observations}

We observed \etacar\ with the \CHANDRA\ X-ray observatory five times
at 5$-$7~day intervals across the 2009.0 deep X-ray minimum (Table \ref{tbl:obslogs}).
\CHANDRA\ observations during the deep minimum are required to spatially resolve emission 
from the binary from the other nearby emission sources, including time-delayed X-ray emission
reflected from the HN \citep{Corcoran2004}.
We placed \etacar\ on the ACIS-S3 chip, which has good quantum efficiency and energy resolution
between 0.3$-$8~keV.
To minimize photon pile-up and instrumental dead time,
we used the 1/8 window mode with 0.4~sec frame exposures.
We used the analysis software CIAO version~4.3, which employs
the energy-dependent sub-pixel event-repositioning (EDSER) algorithm.
Following \citet{Hamaguchi2007b},
individual \CHANDRA\  observations are designated CXO,
subscripted with the year, month and day of the observation.

Eta Carinae was relatively bright in the first observation, CXO$_{090110}$,
a few days before the onset of the deep X-ray minimum, and suffered mild photon pile-up of $\sim$9.7 \%.
During subsequent observations, the photon count rate was sufficiently low that pile-up was unimportant.

\section{Results}
\label{sec:result}

During each observation, \CHANDRA\ detected a point-like source centered on the star (Figure~\ref{fig:obsimg}).
As seen in the 2003.5 deep minimum image \citep{Corcoran2004},
the central point source (CPS) is embedded in an extended 18\ARCSEC\ hard X-ray source,
which is the X-ray emission produced by the WWC, reflected from the bipolar lobes of the HN \citep{Corcoran2004}.
These hard X-ray sources are surrounded by soft X-ray emission extending to 1\ARCMIN\ \citep{Seward2001}.
The faint diffuse structure of the soft X-ray emission is seen clearly,
thanks to the good sensitivity of the ACIS-S imaging in the soft band.

We derived absolute coordinates of the CPS in each observation.
With the source detection software {\tt wavdetect} in the CIAO package,
we measured the CPS position along with other detected X-ray sources above 4$\sigma$ significance in the \FOV\footnote{For this measurement, 
we use data calibrated with the CIAO version 4.2 and 
used the SER method for the sub-pixel algorithm \citep{Li2003,Li2004,Tsunemi2001}.}.
We cross-correlated the detected sources except for the CPS with the X-ray sources identified 
in the Chandra Carina Complex Project \citep{Townsley2011b} and shifted the coordinate frame of 
each observation to match them.
The \FOV\ shift was $\sim$0.33\ARCSEC\ on average.
All the CPS coordinates after this adjustment fell within 0.2\ARCSEC\ from the position of \etacar,
which is within the accuracy of the \CHANDRA\ positional determination.

We also measured the CPS source size for each observation with the CIAO tool {\tt srcextent}.
The CPS size does not change significantly between observations 
at $\sim$0.50\ARCSEC\ $\pm$0.01\ARCSEC\ (1$\sigma$),
and is slightly larger than the \PSF\ images at $\sim$0.44\ARCSEC$\pm$0.04\ARCSEC\ (1$\sigma$),
produced with MARX\footnote{http://space.mit.edu/CXC/MARX/} simulations for similar observing conditions.
Based on this, the CPS could be extended up to $\sim$0.2\ARCSEC ($\sim$460~AU at $d$ = 2.3~kpc).

We extracted X-ray light curves and spectra within 1.5\ARCSEC\ from the CPS  peak, which include more than
95\% of the total point-source flux.
We obtained background within 2\ARCMIN\ from \etacar, excluding areas with X-ray emission associated 
with \etacar\ ($\lesssim$40\ARCSEC), X-ray sources detected at above the 4$\sigma$ level and areas of smaller exposure
due to the satellite dithering motion.
We produced light curves with 1~ksec binning from individual observations;
these did not show any significant variations.

We measured the CPS fluxes for individual observations between 3--8~keV from fits to their spectra by 
an empirical model (bremsstrahlung emission plus 6 Gaussian lines, multiplied by photoelectric absorption) with XSPEC \citep{Arnaud1996}.
We also convolved the {\tt pileup} model 
for a fit to the CXO$_{090110}$ spectrum to account for photon pile-up.
The top panel of Figure~\ref{fig:obstiming} plots the CPS flux variation.
We also added the CPS fluxes measured with \XMM, \SUZAKU\ \citep{Mitsuda2007}, \SWIFT\ \citep{Burrows2005}, and \CHANDRA\ HETGS observations
between the orbital phases 0.97$-$1.06, excluding the \SUZAKU\ observation at the middle of the deep X-ray minimum,
for which the unresolved X-ray contamination of the HN and surrounding sources is not easy to estimate.
For comparison to the earlier published papers, we also overlay the \RXTE/PCA light curves in the 2$-$10~keV band.
The \RXTE\ \FOV\ of 1\DEGREE\ (FWHM) includes many X-ray sources in the field, among which
the brightest hard X-ray source is the Wolf-Rayet (WR) binary, WR25.
We subtracted the estimated contribution of WR 25 (Pollock et al. in preparation) corrected for the collimator response  
($\sim$96\% at the location of WR 25) from the \RXTE\ light curve.
We also corrected the \RXTE\ light curve of the CPS for X-ray emission from weak sources in the \RXTE\ \FOV, 
which we estimated from our \CHANDRA\ deep-minimum observations.
We then converted the \RXTE\ count rates to the energy flux assuming emission from a \KT = 4.5~keV plasma
suffering absorption of \NH = 5$\times$10$^{22}$~\UNITNH, which is appropriate for the absorption through the HN \citep{Hamaguchi2007b}.
The \RXTE\ light curves nicely match measurements by the other X-ray imaging observatories.

\begin{figure*}[hdtp]
\epsscale{1.8}
\plotone{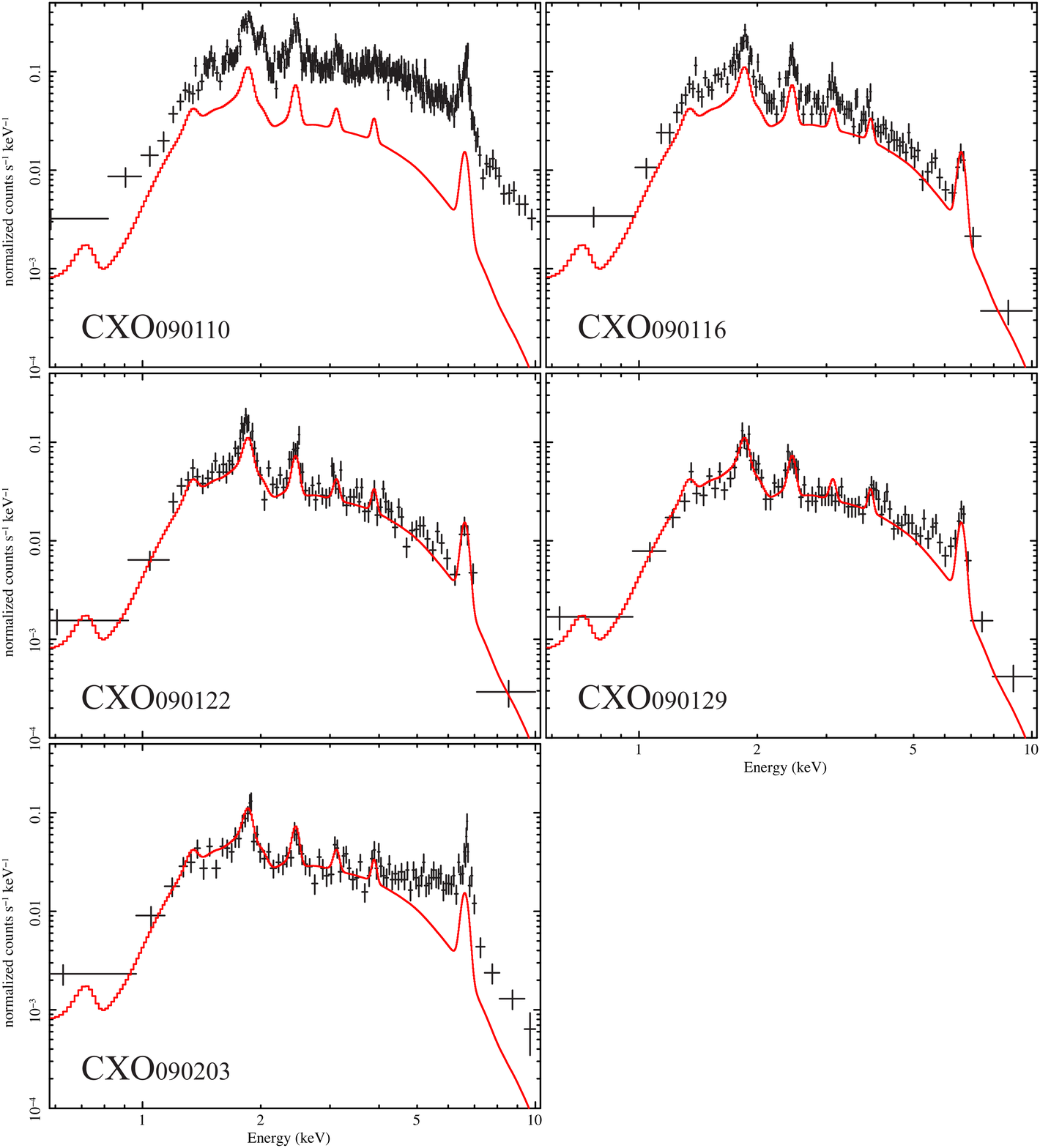}
\caption{
Individual ACIS-S spectra of {\etacar}'s CPS in 2009.
The solid grey (red in the online version) line in each panel shows the lowest observed flux (i.e. the CCE component) measured from a simultaneous fit of these spectra 
by an empirical model (the detailed fitting method will be described in the following paper).
\label{fig:obsspec}
}
\end{figure*}

The first \CHANDRA\ observation (CXO$_{090110}$) was performed when the X-ray flux was rapidly declining
from maximum to minimum.
From this observation to CXO$_{090116}$,
the X-ray flux between 3$-$8~keV dropped by a factor of 6 and decreased slightly more by CXO$_{090122}$ to its minimum observed
of $\sim$1.9$\times$10$^{-12}$~\UNITFLUX.
It then recovered slightly by the 4th observation and further increased by the 5th observation.

Figure~\ref{fig:obsspec} shows spectra of the CPS from all the \CHANDRA\ observations.
Through these \CHANDRA\ observations, the emission was strong between 1$-$10 keV, 
with multiple emission lines from Fe, Ca, Ar, S, Si and Mg ions.
From CXO$_{090110}$ to CXO$_{090116}$, the flux declined  by a factor of $\sim$3$-$4 in both the hard and soft bands (middle panel of Figure~\ref{fig:obstiming}).
After that, the hard band ($\gtrsim$6~keV) spectrum did not change through CXO$_{090122}$, but then it gradually increased at CXO$_{090203}$.
The soft band spectrum gradually decreased until CXO$_{090129}$ and then stayed at that level.
Thus the hard X-ray emission reached its minimum value between CXO$_{090116}-$CXO$_{090122}$,
while the soft X-ray minimum was not reached until sometime between CXO$_{090122}$ and CXO$_{090129}$.

This result clearly shows that the emission from the CPS at its lowest flux is due to the Central Constant Emission (CCE) component
discovered by \citet{Hamaguchi2007b}.
The CCE probably arises from stable hot plasma bubbles outside of the binary system, which are thermalized by collision of ambient gas with 
the secondary's stellar winds ejected in previous orbital cycles as seen in 3-D SPH models \citep{Russell2013,Madura2013}.
The \CHANDRA\ observations in 2009 now show that this spectrum extends up to 
$\sim$10~keV, as we discuss more fully in an upcoming paper (Hamaguchi et al., in preparation).

\section{Discussion}

X-ray emission from the WWC region is the excess over the CCE contribution,
that is, the excess above the solid lines in Figure~\ref{fig:obsspec}.
Hard X-ray emission from the WWC diminished first just before CXO$_{090116}$ and began to recover by CXO$_{090129}$.
Soft X-ray emission from the WWC gradually faded until sometime between CXO$_{090122}$ and CXO$_{090129}$.
The WWC spectrum in CXO$_{090203}$ showed the strongest soft band cut-off among the 2003 and 2009 observations.
We fit this residual WWC component in CXO$_{090203}$ by an absorbed one temperature plasma model with a Gaussian for 
the fluorescent iron K line, fixing the plasma temperature at 4~keV and assuming the solar elemental abundances for both the plasma 
and cold absorber.
The absorption column density is, then, \NH~$\sim$10$^{24}$~\UNITNH, the highest among the \etacar\ observations.
These results suggest that
the hottest X-ray emitting plasma (\KT~$>$3~keV) is localized near the leading side of the bow-shock apex,
which enters into and emerges from an X-ray thick absorbing region;
the softer emission (\KT~$<$3~keV) 
is mostly produced in an extended region associated with the tail of the bow shock, which succeedingly enters behind the thick primary wind.
In this model, the deep minimum is produced by an eclipse of the WWC plasma.
The \CHANDRA\ data show that the onset and end of the deep minimum, barely seen in the \RXTE\ light curve, are similar among 1998, 2003 and 2009, 
which shows that this event is caused solely by orbital motion.
The flux difference at $\phi_{\rm X}\sim1.009$ between 2003 and 2009 (the green and black filled circles in the top panel of Figure~\ref{fig:obstiming}) may be caused by a change in the CCE spectrum
or in the blocking material or both.
A detailed discussion of this point will be discussed in an upcoming paper.

Since the WWC apex is located between the primary and secondary stars,
the deepest part of the X-ray minimum marks the time of superior conjunction.
The exact timing of the deepest eclipse is not easy to estimate
because, near periastron, the secondary star moves rapidly relative to the primary;
the size, relative location and intrinsic luminosity of the WWC plasma change quickly.
Nevertheless, 
linear extrapolations of the hard band (4$-$8~keV)
light curve before and after the hard X-ray minimum (see the middle panel of Figure~\ref{fig:obstiming})
suggests that the deepest eclipse of the WWC apex
occurs between Jan 12 and Jan 28, so that superior conjunction occurs within this timespan.
The onset of the deep minimum, Jan 12, was $\sim$4 days earlier than the predicted using the ephemeris in \citet{Corcoran2005}.
This may suggest that the orbital period is a few days shorter than 2024 days, as suggested by \citet{Damineli2008}.

The shallow minimum was not clearly seen in 2009, 
but there may be a brief shallow minimum interval seen by \SUZAKU\ at $\phi_{\rm X} =$1.015 in 2009.
The strong variation in the X-ray light curves between 2003 and 2009, possibly due to the wind momentum change \citep{Corcoran2010},
occurred during the shallow minimum.
This suggests that the low X-ray activity seen during the shallow minimum is caused by the collapse of the WWC activity around periastron.
The X-ray light curve in 2003 clearly shows a remaining X-ray activity above the CCE flux level during the shallow minimum.
Since the observed hard X-ray emission above 6~keV did not change significantly during the shallow minimum in 2003 \citep[see Figure~4 in][]{Hamaguchi2007b},
this hard X-ray emission must not depend on the orbital motion of the companion.
This residual emission perhaps arises from shocked gas produced by the collision of the companion's wind with the "wall" of primary wind 
extending behind the companion near periastron.
The complete X-ray minimum seen by \RXTE\ is thus the result of a hybrid mechanism: an eclipse of the WWC plasma during the deep minimum
followed by a collapse of the WWC activity seen during the shallow minimum.

The collapse of the WWC shock is very probably 
related to the extreme environment near periastron, e.g., high density 
(which could produce a dramatic increase in the cooling rate)
or strong UV radiation (which could radiatively inhibit the companion's wind).
This suggests that periastron passage occurs sometime during the decay of the intrinsic X-ray activity.
Since the WWC emission is not seen directly during the deep minimum,
the wind collapse may have started earlier than the shallow minimum,
which is defined by transition of the observed X-ray light curve:
the onset of the wind collapse can be as early as the start of the deep minimum.
The \RXTE\ light curve starts to recover by Feb. 18, so that
the WWC collapse, and periastron passage, probably occurs between $\phi_{\rm X} =$0.998$-$1.016.
In this solution, periastron comes around superior conjunction, i.e.,
our view from the argument of periapsis ($\omega$) around 270\DEGREE.
This is consistent with a 3-D hydrodynamic simulation of the X-ray light curve \citep{Okazaki2008,Parkin2011,Russell2013}.

The hard X-ray emission from the WWC region dropped concurrently with the He~II $\lambda 4687$\AA\ emission line
\citep[][see middle and bottom panels of Figure~\ref{fig:obstiming}]{Teodoro2012}.
The HeII $\lambda 4687$\AA\ emission line dropped on Jan 15$\pm$1 and recovered on Jan 26$\pm$2,
which agrees quite well with the hard X-ray drop between $\sim$Jan 12 and $\sim$Jan 28.
This result indicates that the He~II $\lambda 4687$\AA\ line is produced near the WWC apex.
The HeII $\lambda 4687$\AA\ line requires recombination of He$^{++}$ ions \citep{Steiner2004,Martin2006,Teodoro2012},
which in principle can be supplied by shocks at the WWC apex.

The intrinsic WWC activity is still strong at the beginning of the deep minimum,
so that the occulter must be sufficiently optically thick to block WWC X-rays up to $\sim$10~keV
(optical depth $\tau\gtrsim1$ below 10~keV when \NH~$\gtrsim$5$\times$10$^{23}$~\UNITNH).
One candidate for the occulter is the photosphere of \etacar\ A itself.
Within the expected ranges of orbital parameters ($e\sim$0.85$-$0.95, $\omega\sim$240$-$285\DEGREE, inclination $\sim$35$-$55\DEGREE,
wind momentum ratio $\sim$0.24),
the occulter has to be 180$-$320~\UNITSOLARRADIUS\ to block the hard X-ray emission for $\Delta\phi_{\rm X} \sim$7.8$\times$10$^{-3}$,
the phase interval of the hard X-ray drop during the deep minimum.
This is significantly larger than the current best estimate of the primary stellar radius ($\sim$60\UNITSOLARRADIUS),
which has, however, large uncertainty \citep{Hillier2001}.
Another candidate is the primary wind.
When we assume a smooth wind of the solar abundance matter flowing with the velocity of $\sim$420~\UNITVEL\ and 
the mass loss rate of $\sim$8.5$\times$10$^{-4}$~\UNITSOLARMASSYEAR\ \citep{Groh2012},
the hydrogen column density to the WWC apex can be as high as $\approx$8$\times$10$^{24}$~\UNITNH\ at the impact parameter of 500~\UNITSOLARRADIUS,
which is enough to block X-ray emission at 10 keV.
The X-ray spectrum of CXO$_{090203}$ shows an absorption column to the WWC emission of \NH~$\sim$10$^{24}$~\UNITNH (assuming solar abundance).
The absorption column to the WWC plasma would be much higher than this \NH\ in the middle of the deep minimum.

\section{Summary}

\CHANDRA\ observations during the X-ray minimum in 2009 have for the first time given us a detailed picture of 
the behavior of X-ray emission from the WWC between the primary and secondary stars in the extremely 
massive, long-period, highly eccentric binary \etacar.
We interpret the observed spectrum variations as driven by orbital motion: an initial occultation of the apex of the wind-wind bow shock 
followed by a disruption of the shock due to increase in density and/or UV radiation near periastron passage.
Using these data, we constrain superior conjunction to occur between X-ray phase 0.998$-$1.006, and periastron passage to occur in the 
X-ray phase interval 0.998$-$1.016.
The large opacity at high energies during the ``deep minimum" suggests occultation by the thick inner wind or
by the stellar photosphere of the primary LBV itself.
If the latter, then the periastron distance is comparable to the radius of the primary star's photosphere.

\acknowledgments

This work was performed while K.H. was supported by the NASA's Astrobiology Institute (RTOP 344-53-51) 
to the Goddard Center for Astrobiology (Michael J. Mumma, P. I.).
This research has made use of data obtained from the High Energy Astrophysics Science Archive
Research Center (HEASARC), provided by NASA's Goddard Space Flight Center.
We appreciate the \CHANDRA\ operation scientist Scott J. Wolk for important advices on enhancing efficiency of the \CHANDRA\ observations and 
useful comments on the paper draft by David B. Henley and Kris Davidson.

{\it Facilities}: \facility{CXO (ACIS-S, HETGS)}, \facility{XMM}, \facility{Suzaku}, \facility{Swift}, \facility{RXTE}


\end{document}